\documentclass{article}
\pdfoutput=1
\usepackage{amsmath,amssymb,epsfig,color}

\def\de#1#2{{\rm d}^{#1}  #2 \, }

\def\sqr#1#2{{\vcenter{\vbox{\hrule height.#2pt
 \hbox{\vrule width.#2pt height#1pt \kern#1pt
 \vrule width.#2pt}\hrule height.#2pt}}}}

\newcommand{\beq}{\begin{equation}}
\newcommand{\eeq}{\end{equation}}
\newcommand{\alphabar}{\bar{\alpha}_s}
\newcommand{\betabar}{\bar{\beta_0}}

\def\aa1{\phi}
\def\cc1{\psi}

\newcommand{\abs}[1]{\left| #1 \right|} 
\newcommand{\norm}[1]{\lvert #1 \rvert} 

\newcommand{\drv}{{\rm d}}

\title{A few topics in BFKL phenomenology at hadron colliders\label{ch1} } 
\author{Agust{\' \i}n Sabio Vera\\ 
\\
CERN, Theoretical Physics Department, Geneva, Switzerland.\\
\\
Instituto de F{\' \i}sica Te{\' o}rica UAM/CSIC, Nicol{\'a}s Cabrera 15\\ 
\& U. Aut{\' o}noma de Madrid, E-28049 Madrid, Spain.} 

\begin{document} 

\maketitle 

\begin{abstract}
This is a personal recollection of several results involving the phenomenological study of the multi-Regge limit of scattering amplitudes. None of them would have been possible without the encouragement and constant support from Lev Nikolaevich Lipatov.
\end{abstract}

\section{Lev in Madrid}

The first time I met Lev N. Lipatov was at a Low x Physics workshop at the end of June of 2001, in Krakow, Poland\footnote{http://th-www.if.uj.edu.pl/low\_x}. At the time I was a postdoctoral fellow at the Cavendish Laboratory, in Cambridge, and I thought I knew everything about the high energy limit of Quantum Chromodynamics (QCD). I had been introduced to the concept of the Pomeron in Regge theory by Sandy Donnachie, and had worked during my PhD thesis in Manchester with Jeff Forshaw and Douglas Ross on the understanding of the fundamentals of the Balitsky-Fadin-Kuraev-Lipatov (BFKL)~\cite{Lipatov:1985uk,Balitsky:1978ic,Kuraev:1977fs,Kuraev:1976ge,Lipatov:1976zz,Fadin:1975cb}  equation at next-to-leading order. Things did not go as planned the day of my talk at that conference because, right after my presentation, Lev stood up and said out loud that all what I had said was completely wrong. Suddenly it felt like a huge black hole opened up under my feet and I got sucked down into it. 

After that very rough beginning we became good friends and collaborators for many years. The last time we met and worked together was at the beginning of August 2017, when Lev and his wife Elvira came to Cambridge where I was spending the summer at DAMTP.  Already from our first discussions it was clear he was a very special character, full of energy, and an amazing mind. He had an incredible knowledge of all the important works from the old times, before Quantum Field Theory, which he was able to blend with the modern developments in a way I have not witnessed in any other researcher. When working with him you had to be ready to be carried to any uncharted territory. You could start talking about radiative corrections in QED and end up with a blackboard full of results on quantum integrability and tachyon scattering in string theory.  When discussing with Lev you always had the impression that something great was about to happen, that a completely new and brilliant idea was about to arise, and this would immediately motivate you to work on the problem at hand as if there were no tomorrow.

We used to meet regularly mainly at four locations: in Hamburg, hosted at DESY by Jochen Bartels; in Geneva, at the CERN Theory Group (then Unit and now Department); in Gatchina, at the Petersburg Nuclear Physics Institute, where he was Director of the Theory Division; and in Madrid, at Aut{\'o}noma university and the Instituto de F{\'\i}sica Te{\' o}rica, where he was Severo Ochoa (SO(IFT)) Distinguished Professor. During their many visits Lev and Elvira lived in my home town, Alcal{\'a} de Henares, and we used to spend endless hours trying to push our projects forward. Funnily  enough, we did not publish any papers on QCD but more on supersymmetric and gravitational theories. Nevertheless his encouragement and insight were crucial for the development of the several results which will be described in the following. As a disclaimer, allow me to indicate that this is not a comprehensive review of the subject, it is just a personal recollection of results on the phenomenology of the BFKL formalism which gather together many of the discussions we had. 

\section{Where are the Reggeon fields?}

Let us start by motivating the experimental search of BFKL effects at colliders. Right now we are at an interesting period to investigate this subject because we have 
good control of the theoretical calculations and the Large Hadron Collider (LHC) is operating at full power. This should allow us to determine the region of applicability of asymptotic calculations of scattering amplitudes  in the high energy limit. 

The study of scattering processes in the high energy limit of QCD offers invaluable information that goes beyond the perturbative aspects of the theory. It touches important issues such as factorization between soft and hard physics and it challenges fixed order calculations. At high energies, the convergence of the perturbative expansion, truncated at a certain order in the strong coupling $\alpha_s$, is not \textit{a priori} guaranteed. This is because large logarithms 
of the  center-of-mass energy  squared, $s$, 
appear in Feynman diagrams and it is needed to resum them to all orders. 
A powerful approach to perform this resummation is the BFKL framework initially developed at the leading logarithmic (LL) approximation~\cite{Balitsky:1978ic,Kuraev:1977fs,Kuraev:1976ge,Lipatov:1976zz,Fadin:1975cb,Lipatov:1985uk}, where terms of the form $(\alpha_s \log(s))^n$ were resummed. In order to improve those results, the next-to-leading logarithmic (NLL) approximation  corrections to the BFKL kernel were calculated~\cite{Fadin:1998py,Ciafaloni:1998gs}, where also terms that behave like $\alpha_s (\alpha_s \log s)^n$ were taken into account.  It was soon realised, however,  that at NLL the positiveness of cross sections was not always ensured. This is due to the presence of large collinear logarithms that need extra treatment, a step that led to the so-called collinearly improved kernel~\cite{Salam:1998tj,Ciafaloni:2003ek} allowing for more robust phenomenological studies. Obviously, an important question for collider phenomenology is gauging reliably at which energies  the BFKL dynamics becomes relevant and cannot be ignored. 

What are the observables where this formalism should apply? In simple terms, in a scattering amplitude we identify and calculate those contributions at each order of perturbation theory which have the largest numerical value when $s$ is big.  This shares a common nature with other resummation schemes where a class of terms has to be calculated to all orders. What makes this formalism different is that it is not just a way to calculate some observables more precisely, it also shows that at high energies new degrees of freedom, reggeized particles, play a dominant role. These reggeons are universal in the sense that they drive a large variety of cross sections with very different initial and final state configurations, present in lepton-lepton, lepton-hadron and hadron-hadron colliders. Remarkably, this effective picture also holds for electroweak, supersymmetric and gravitational interactions. 

The reggeization picture~\cite{Fadin:1998fv}, at LL and NLL accuracies, is robust and has been cross-checked by many different methods, including elegant techniques in string theory~\cite{Bajnok:2008qj}. It is in this linear NLL approximation where we have the opportunity to test the applicability of these emergent degrees of freedom to the LHC phenomenology program since $s$ is large enough to justify the use of reggeized propagators but not so large as to necessarily have to introduce non-linear corrections which go beyond NLL. It is then a timely matter to find out where the window of applicability of the formalism lies. This means to identify observables where this approach is distinct, {\it i.e.}, quantities where it fits the data and other possible approaches (fixed order or other resummations implemented in general Monte Carlo event generators) fail. 

\begin{figure}
\vspace{-1.cm}
  \centering
  \includegraphics[width=11.cm,angle=0]{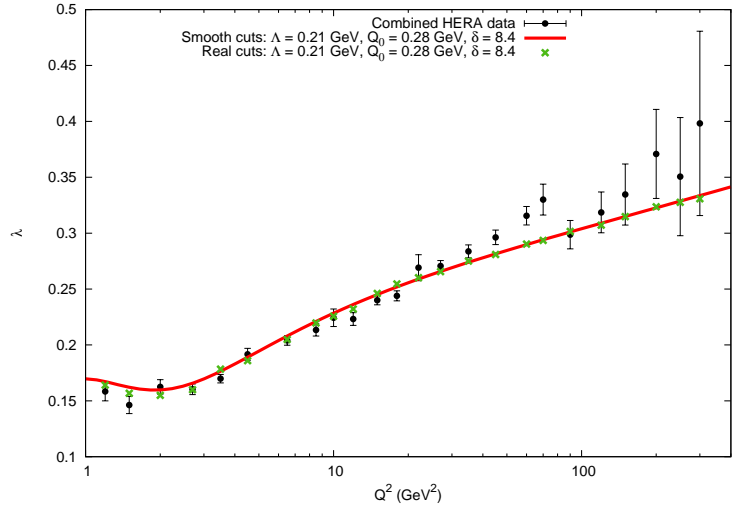}
  \vspace{-1.cm}
  \caption{{\small BFKL calculation for $\lambda(Q^2)$ in the parametrization of the structure function $F_2 (x,Q^2) = c (Q^2) x^{-\lambda(Q^2)}$ at NLL with collinear improvements and a model for the proton impact factor of the form $\simeq (p^2/Q_0^2)^\delta e^{-p^2/Q_0^2}$}, for values of $x< 0.01$.}
  \label{F2FL}
\end{figure}

The attempts to search for BFKL effects have had the difficulty of being applied at  too low energies or rapidity differences in the final state as to be conclusive. A further drawback is that they dealt with too inclusive observables as to be able to claim that the cross section under study could be described by BFKL dynamics and nothing else. A paradigmatic example is the description of the growth of the proton structure functions at low values of Bjorken $x$. Indeed it is possible to get a good fit of the combined HERA data for the $F_2$ and 
$F_L$ structure functions in the small $x$ region with a NLL BFKL calculation. As an example,  see Fig.~\ref{F2FL}, taken from~\cite{Hentschinski:2012kr,Hentschinski:2013id}. 
However, it is  possible to fit these data with other approaches as well. Moreover, there is model dependence in the BFKL calculation itself, which includes three free parameters for the coupling of the gluon ladder to the hadron, a freezing of the coupling in the infrared and the use of a collinear resummation together with a particular renormalization scheme which allows to reach low values of $Q^2$ in the fit. All of this pushes us to try to find other, less inclusive, observables to test small $x$ resummations in a more distinct manner.

\section{More exclusive observables}

The LHC energies are higher than those at the Tevatron or HERA and there is enough statistics to allow for the study of more exclusive quantities, fine-tuned with stringent cuts as to be driven by multi-Regge kinematics, which is the underlying  principle behind the BFKL approach.  The experimental studies should not only consider  the usual ``growth with energy" signal, associated to the exchange of a hard Pomeron, but also other footprints related to energy flows and azimuthal angle dependences. We will motivate this below.

\subsection{Tagging of forward jets hides the growth with energy}

It is only possible to remain in the multi-Regge kinematics if the impact factors associated to each of the incoming protons are strongly peaked at a single hard scale, similar at both hadrons. It turns out that the hard scale is typically associated to the production of forward jets calculated in collinear factorization. This is known as  Mueller-Navelet (MN) jet production~\cite{Mueller:1986ey}. MN jets have  inclusive final states where two jets with transverse momenta of similar sizes, $k_{A,B}$, are tagged when they are widely separated in rapidity $Y$. 

\begin{figure}
\centering
\includegraphics[scale=0.4]{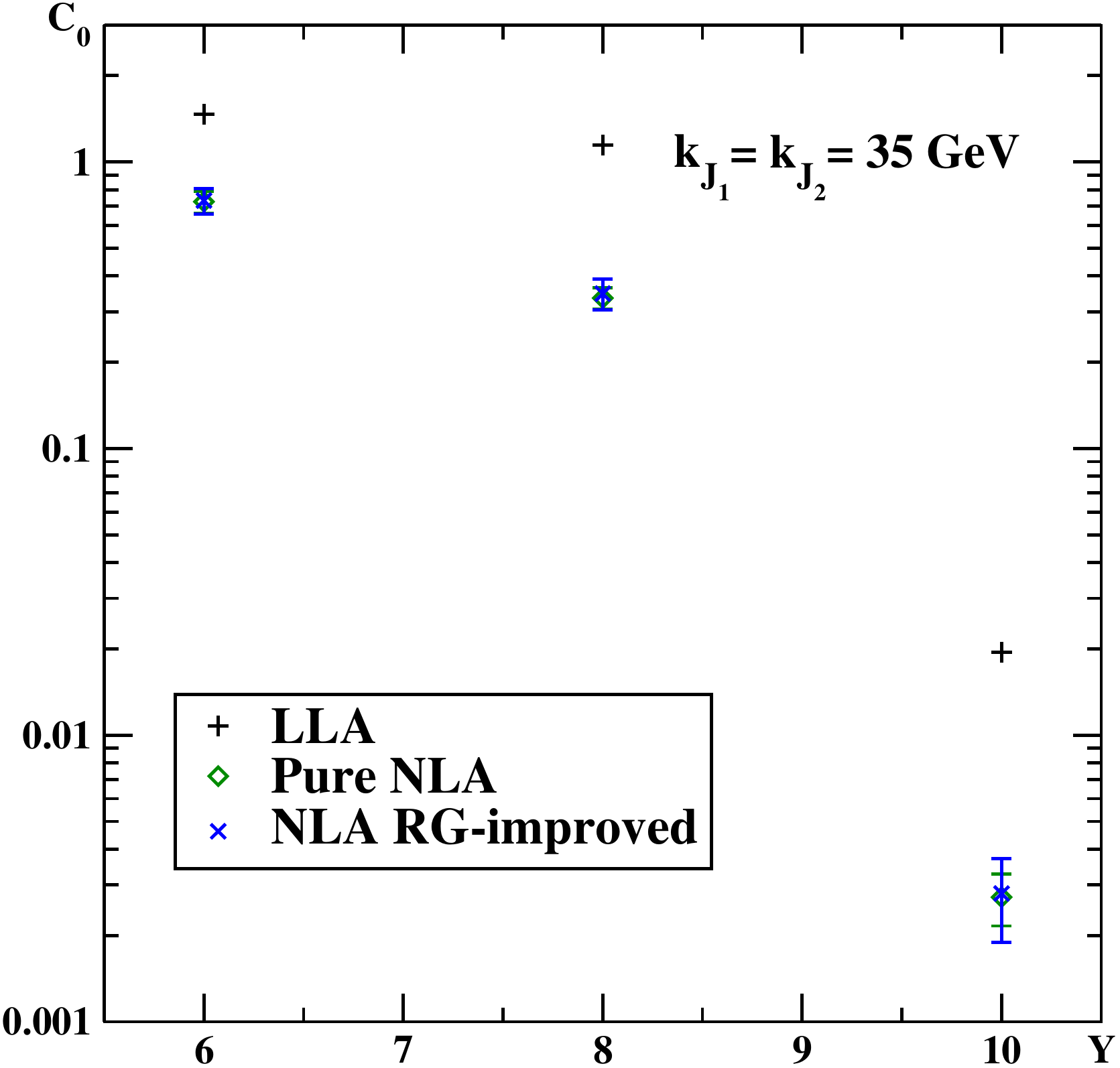}
\caption[]{$Y$ dependence of MN cross section $C_0$ for $|\vec k_{J_{1,2}}|=35$ GeV at $\sqrt s=14$ TeV.}
\label{MNC0}
\end{figure}

The further the MN jets are separated in rapidity, the deeper into the $x \to 1$ limit of the collinear parton distribution functions we enter. 
This is  a region characterized by a drop of the cross section due to energy momentum conservation. This hides the BFKL effects in terms of growth of the cross section as can be seen in Fig.~\ref{MNC0} (obtained in~\cite{Caporale:2013uva}) where the cross section for MN jets with the same lower cut in the $p_T$ of the tagged jets is investigated as a function of their rapidity difference for a LHC run at $\sqrt{s} =  14$ TeV. The cross section decreases as $Y$ increases. 

\subsection{Ratios of azimuthal correlations as a clean signal}

A different signal~\cite{DelDuca:1993mn,Stirling:1994he} which is sensitive to the wealth of the BFKL formalism is to look at the dependence on the relative 
azimuthal angle, $\phi$, between the two MN jets, in the form of $n$-th moments.  The emission of multiple mini-jets manifests as a fast decrease of $< \hspace{-.12cm}\cos{(n \phi)} \hspace{-.12cm}>$ with $Y$. However, even these differential distributions suffer from a large influence of the collinear region. This is due to the fact that $< \hspace{-.12cm}\cos{(n \phi)} \hspace{-.12cm}> \simeq \exp{(\alpha_s Y (\chi_n (1/2) - \chi_0(1/2)))}$, where 
$\chi_n(\gamma)$ is, in Mellin space, the $n$-th Fourier component of the BFKL kernel. $\gamma \simeq 1/2$ dominates the cross section. 

It turns out that the $n=0$ component is very sensitive to collinear dynamics well  beyond the original multi-Regge kinematics. Even though it is possible to include these collinear terms by modifying the original BFKL equation, we believe it is more important to first find the region of applicability of the original formalism by using observables which are far less sensitive to this collinear ``contamination". It is only following this philosophy that we will be able to find ``distinct" BFKL observables. A relevant step in this direction was taken in~\cite{Vera:2006un,Vera:2007kn} where instead of $< \hspace{-.12cm}\cos{(n \phi)} \hspace{-.12cm}>$ it was proposed to remove the $n=0$ dependence by studying ``conformal ratios"\footnote{They capture the $SL(2,C)$ nature of the effective theory of QCD at high energies. When these ratios are calculated at NLL accuracy in the $N=4$ supersymmetric Yang-Mills model, with four-dimensional conformal invariance, they are in agreement with those obtained in QCD using the Brodsky-Lepage-Mackenzie (BLM) scale-fixing procedure in momentum-substraction (MOM) renormalization scheme (see~\cite{Angioni:2011wj}).} 
${\cal C}_{m,n} = < \hspace{-.12cm}\cos{(m \phi)} \hspace{-.12cm}> / < \hspace{-.12cm}\cos{(n \phi)} \hspace{-.12cm}>$ which 
behave like $\simeq \exp{(\alpha_s Y (\chi_m (1/2) - \chi_n (1/2)))}$. It is important to note that the BFKL kernel for $n \neq 0$ is asymptotically insensitive to collinear regions. In~\cite{Vera:2006un,Vera:2007kn} it was proven that these new ratios are very stable under radiative corrections with the LL result (even with running of the coupling) being very similar to the NLL prediction. 

\begin{figure}
\begin{center}
\vspace{-0.5cm}
\hspace{-.7cm}\includegraphics[height=2.4in]{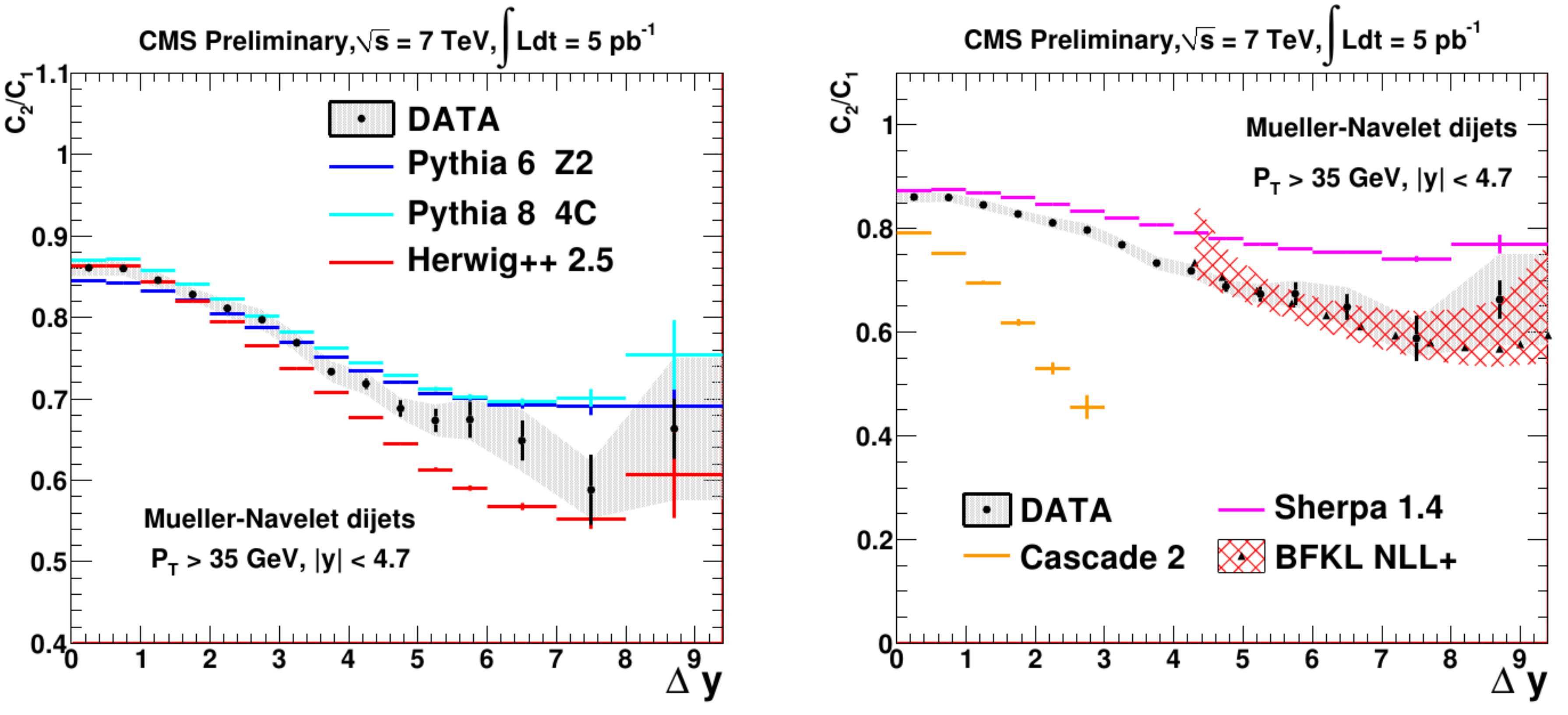}\\
\vspace{-0.6cm}
\end{center}
\caption{The ratio of average cosines ${\cal C}_{2,1} = C_2/C_1$ in bins of  $\Delta y = Y$, compared to various Monte Carlo models.}
\label{C21vsMC}
\end{figure}

With the advent of the LHC it was soon realized that the NLL predictions for these ratios, including NLO forward jet vertices, were in agreement with the data. Furthermore, these observables are so fine-tuned  to the multi-Regge limit that it was difficult for other approaches to describe them.
This could be seen in, {\it e.g.}, the studies presented in~\cite{Ciesielski:2014dfa}, see Fig.~\ref{C21vsMC}, where only a BFKL analysis at NLL is able to fit the large $Y$ tail of the Mueller-Navelet ``conformal ratios" proposed in~\cite{Vera:2006un,Vera:2007kn}. In ~\cite{CMS:2013eda,Khachatryan:2016udy} a comparison between data and theory for a number of MN azimuthal ratios is shown. More recent studies can be found in~\cite{Ducloue:2013bva,Colferai:2010wu}. 

A generalization of the azimuthal ratios has been proposed for processes that have three~\cite{Caporale:2015vya,Caporale:2016soq}
and four final state jets~\cite{Caporale:2015int,Caporale:2016xku}. These are  special MN cases since 
the outermost jets still have a large rapidity distance and other tagged jets are central.

\section{The Monte Carlo event generator {\tt BFKLex}}

For future developments, two main ingredients are needed: new process dependent NLO impact factors  and Monte Carlo techniques to control the gluon Green function and to extract its physical content in the most differential form. The former are mandatory to fairly test the theory and correctly control the dependence on the scales appearing in the calculations (running of the coupling and energy scale separating the universal gluon radiation in central regions of rapidity and that stemming from the fragmentation regions). The latter are needed in order to effectively generate differential distributions which are difficult to obtain analytically.

Let us present several observables characterizing multi-jet configurations event by event (averages of  transverse momentum, azimuthal angle and ratios of jet rapidities). They can be studied using the Monte Carlo event generator {\tt BFKLex}, developed with Grigorios Chachamis,  where higher-order collinear corrections are implemented together with the NLO kernel. 

\begin{figure}
\begin{center}
\includegraphics[height=7cm]{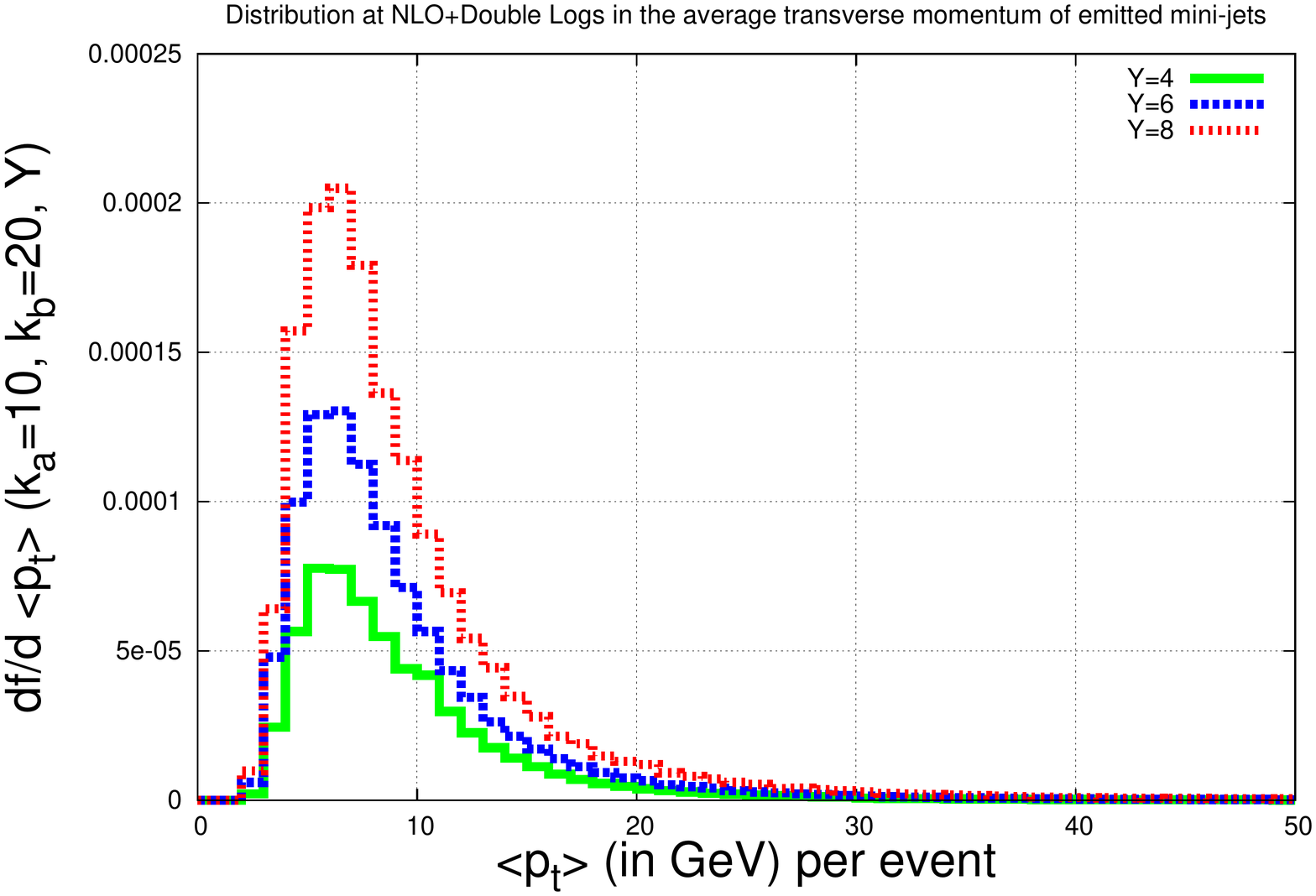}\\
\vspace{-1.cm}
\includegraphics[height=7cm]{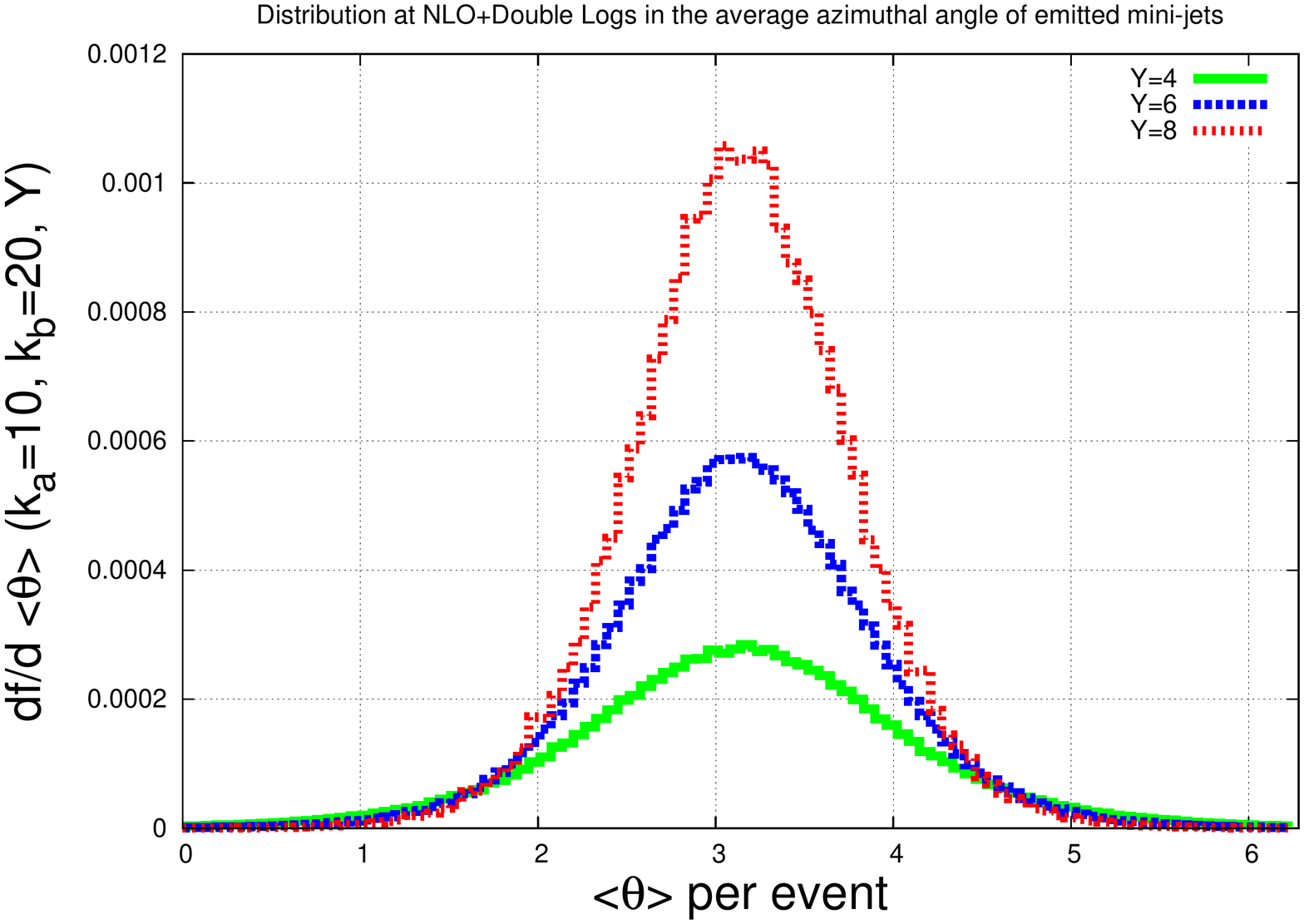}\\
\vspace{-1.cm}
\includegraphics[height=7cm]{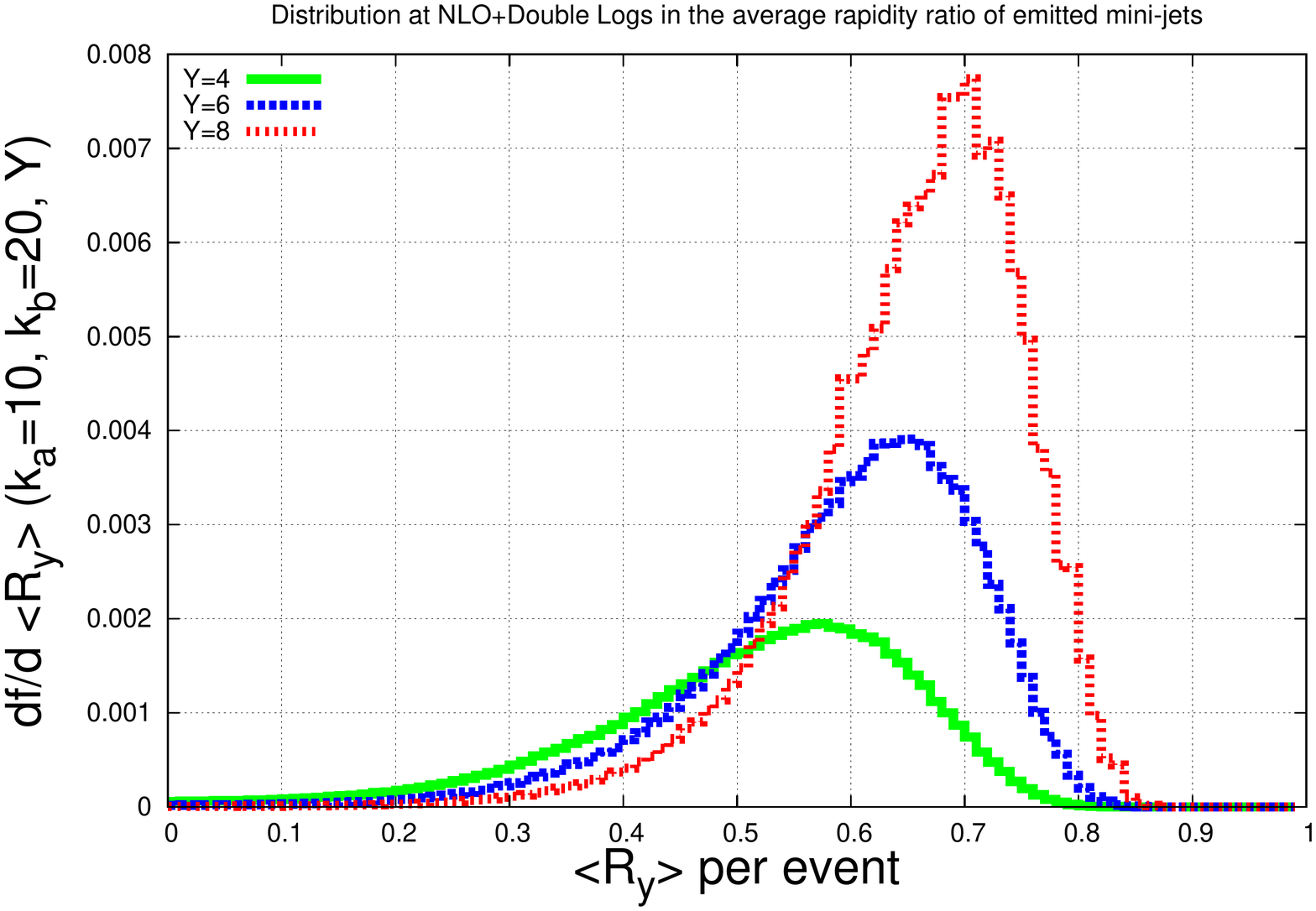}
\end{center}
\vspace{-.4cm}
\caption{High-multiplicity distributions generated with {\tt BFKLex}.}
\label{Plots}
\end{figure}

The BFKL formalism allows for the investigation of high-multiplicity final-states even if the scattering energy is not large. If we require two tagged forward/backward jets in the final state the cross sections can be written in the form
\begin{eqnarray}
\sigma (Q_1,Q_2,Y) = \int d^2 \vec{k}_A d^2 \vec{k}_B \, {\phi_A(Q_1,\vec{k}_a) \, 
\phi_B(Q_2,\vec{k}_b)} \, {f (\vec{k}_a,\vec{k}_b,Y)}.
\end{eqnarray}
The impact factors $\phi_{A,B}$ depend on hard scales $Q_{1,2}$.  The gluon Green function $f$ depends on the transverse momenta $\vec{k}_{a,b}$ and the center-of-mass energy in the scattering or, alternatively, on the rapidity difference between the two tagged jets. At LL and NLL $f$ admits the iterative representation~\cite{Schmidt:1996fg,Andersen:2003an,Andersen:2003wy}
\begin{eqnarray}
f &=& e^{\omega \left(\vec{k}_A\right) Y}  \Bigg\{\delta^{(2)} \left(\vec{k}_A-\vec{k}_B\right) + \sum_{n=1}^\infty \prod_{i=1}^n \frac{\alpha_s N_c}{\pi}  \int d^2 \vec{k}_i  
\frac{\theta\left(k_i^2-\lambda^2\right)}{\pi k_i^2} \nonumber\\
&\times& \int_0^{y_{i-1}} \hspace{-.3cm}d y_i e^{\left(\omega \left(\vec{k}_A+\sum_{l=1}^i \vec{k}_l\right) -\omega \left(\vec{k}_A+\sum_{l=1}^{i-1} \vec{k}_l\right)\right) y_i} \delta^{(2)} \hspace{-.16cm}
\left(\vec{k}_A+ \sum_{l=1}^n \vec{k}_l - \vec{k}_B\right) \hspace{-.2cm}\Bigg\}, 
\label{simpleIter}
 \end{eqnarray}
with $\omega \left(\vec{q}\right)$ being the gluon Regge trajectory and $\lambda$ an infrared cut-off. {\tt BFKLex} implements this iteration and has been used for collider phenomenology and other formal studies~\cite{Chachamis:2013rca}. As we have discussed, the BFKL formalism is quite sensitive to collinear configurations. The dominant double-log terms in this region can be resummed~\cite{Salam:1998tj}. In~\cite{Vera:2005jt}, it was shown that this can be done in transverse momentum representation using the following prescription in Eq.~(\ref{simpleIter}):
\begin{eqnarray}
\theta \left(k_i^2-\lambda^2\right) \to \theta \left(k_i^2-\lambda^2\right)  + \sum_{n=1}^\infty 
\frac{\left(-\bar{\alpha}_s\right)^n}{2^n n! (n+1)!} \ln^{2n}{\left(\frac{\vec{k}_A^2}{\left(\vec{k}_A+\vec{k}_i\right)^2}\right)}. 
\label{SumBessel}
\end{eqnarray}
This step improves the convergence of the perturbative expansion and in $\gamma$ representation corresponds to a $J_1$-Bessel function originally found in~\cite{Vera:2006un}.  It has been implemented in {\tt BFKLex}~\cite{Chachamis:2015zzp} to study MN jet production~\cite{Chachamis:2015ico}. Three averages in each event have been investigated: of the modulus of the transverse momentum of the final-state jets, of their azimuthal angle and of the rapidity ratio between subsequent jets:
\begin{eqnarray}
\langle p_t \rangle ~=~ \frac{1}{N} \sum_{i=1}^{N} |k_i|;
\label{eq:observable1} \,\, \,\,\, \,
\langle \theta \rangle ~=~ \frac{1}{N} \sum_{i=1}^{N} \theta_i;
\label{eq:observable2} \, \, \,\,\, \,
\langle {\mathcal R}_y \rangle =~ \frac{1}{N+1}  \sum_{i=1}^{N+1} \frac{y_i}{y_{i-1}}.
\label{eq:observable3}
\end{eqnarray}
Some configurations under study include $k_a = 10$ GeV, $k_b = 20$ GeV and  $y_a-y_b = 4, 6, 8$.  Fig.~\ref{Plots}
shows broad distributions in $\langle p_t \rangle$ with a maximal value at $\langle p_t \rangle \simeq 6$ GeV.  The cross section receives important contributions from jets with large transverse momentum.  The azimuthal angle between the two tagged jets changes randomly event by event. The $\langle \theta \rangle$ per event at which the untagged jets are produced is shown in Fig.~\ref{Plots} together with the mean ratios of rapidities  $\langle {\mathcal R}_y \rangle$.
The distributions have maxima at $\langle {\mathcal R}_y \rangle > 0.5$. As they are quite broad, this means that there are large contributions  from preasymptotic configurations.

\begin{figure}
\begin{center}
\vspace{-0.5cm}
\includegraphics[width=7.5cm]{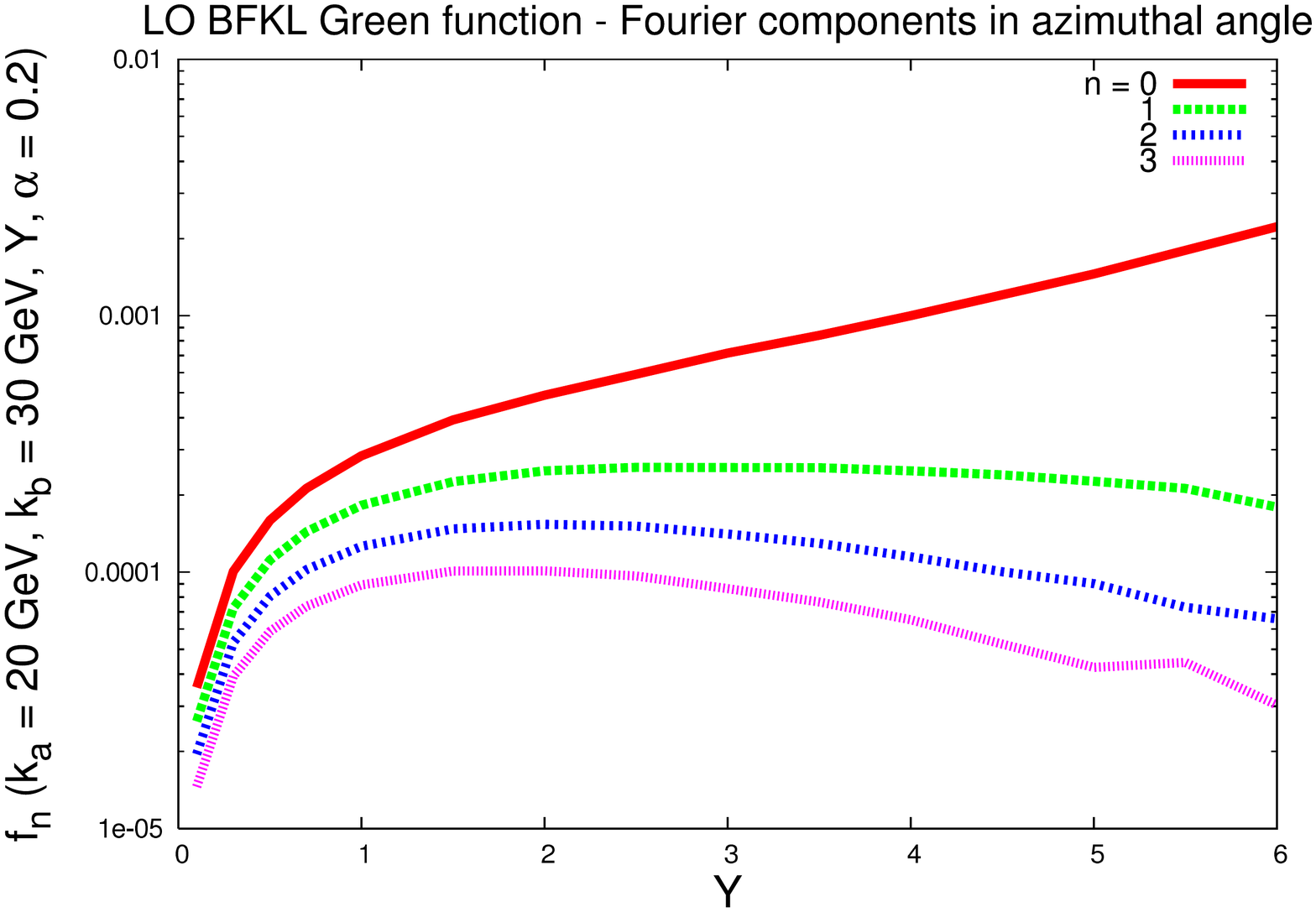}
\includegraphics[width=8.5cm]{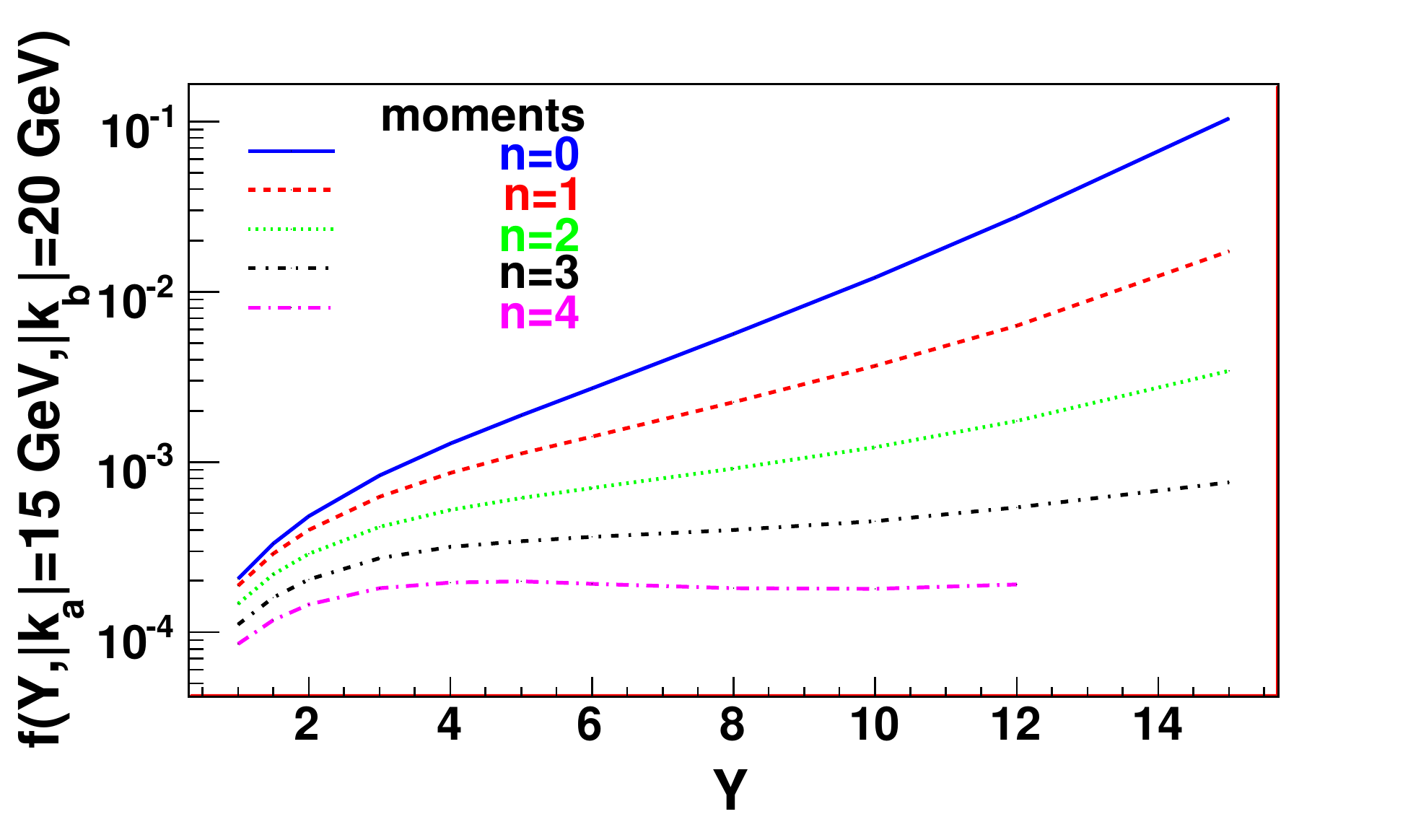}\\
\vspace{-0.6cm}
\end{center}
  \caption{Rapidity dependence of the azimuthal angle Fourier components in the BFKL (top) and CCFM (bottom) Green function.}
  \label{BFKLvsCCFMFourier}
\end{figure}

The value of Monte Carlo studies  can be illustrated with another example related to the azimuthal angles. Let us compare the above mentioned $n$-moments of the BFKL cross section with those obtained from the Catani-Ciafaloni-Fiorani-Marchesini (CCFM)~\cite{Catani:1989yc,Marchesini:1994wr,Ciafaloni:1987ur,Catani:1990gu} approach, which,  in principle, interpolates the large and small $x$ limits of unintegrated gluon densities (we will see that this is only true for the $n=0$ projection). This was studied in~\cite{Chachamis:2011rw} and it is shown here in Fig.~\ref{BFKLvsCCFMFourier}. In the top plot we see the BFKL result and at the bottom the CCFM analysis. We can see that the $n=0$ component has a similar behavior in both cases since it grows with  $Y$. But this is not the case for the $n>0$ components which in the BFKL case decrease with $Y$ and for CCFM increase. This is a fundamental difference between the Regge limit and approaches based on QCD coherence~\cite{Gonzo:2019fai} which deserves to be explored further in order to disentangle BFKL from other dynamics. 

Regarding the impact factors and jet vertices, with Martin Hentschinski, we have developed a method to calculate them based on Lipatov's high energy effective action~\cite{Hentschinski:2011tz,Chachamis:2013hma,Chachamis:2013qca,Chachamis:2013oga}. Not only we have reproduced the NLO forward jet vertex coupled to an octet~\cite{Chachamis:2012gh,Chachamis:2012mw,Chachamis:2012cc}, which is related to forward jet production with associated mini-jet radiation but also when the jet plus proton remnants are attached to a color singlet~\cite{Hentschinski:2014lma,Hentschinski:2014bra,Hentschinski:2014esa}, which is associated to a gap in rapidity without any hadronic activity in the detectors. For these diffractive events we have the ``traditional" problem of the gap survival probability which is difficult to calculate in a reliable manner. {\tt BFKLex} can be very useful in this context of diffractive physics. Further processes at full NLL, like forward production of electroweak bosons, will also require the calculation of NLO impact factors, a problem which can be suitably addressed within the high energy effective action approach. 

It is important to put forward this program since there are uncertainties in the BFKL approach itself which need to be fixed. As an example, it is needed to find the dependence of each proposed observable on the renormalization schemes (the above mentioned conformal ratios were shown to be independent of these choices), but also the correct treatment of the running of the coupling must be addressed in an accurate way (we will discus this further below). Only a fair comparison to experimental data can solve many of these theoretical questions. 

If we control some observables which can only be described by BFKL then it makes sense to introduce corrections in the form of hadronization, non-linearities, collinear radiation or study connections with soft-collinear effective theories, in order to extend their range of applicability beyond the multi-Regge kinematics.

\section{The challenge of the infrared limit}

The structure of the BFKL equation changes in the infrared region. A pressing question is to find the optimal treatment of the running of the coupling. This is one of the many attractive features of this formalism since it offers the possibility to study the transition from perturbative to non-perturbative dynamics in a continuous way. Lipatov~\cite{Lipatov:1985uk} worked on this problem and proved that, if suitable infrared boundary conditions are imposed when treating the running of the coupling, the cut  in the complex angular momentum plane transforms into an infinite series of poles. For him this was important since it connected with Gribov's ideas where high energy scattering could be dominated by Regge poles after all. 

In~\cite{Ross:2016zwl}, together with Douglas Ross, we realized that integrating along a contour off the real axis we find a strong dependence of the intercepts and collinear regions on the choice of the boundary conditions.  We found that the mean transverse scales dominant in the gluon ladder increase, pushing the gluon evolution towards harder scales. If this is the correct treatment of the running of the coupling, it could have interesting phenomenological consequences worth exploring. 

The nutshell of the calculation can be explained using the notation  $t_i \equiv \ln(k_i^2/\Lambda^2_{\mathrm{QCD}})$,  $ \bar{\alpha}_s \ \equiv \  \frac{C_A}{\pi}  \alpha_s$,  $ \bar{\alpha}_s(t) \ = \ \frac{1}{\bar{\beta_0} t} $, and a form of 
 running the coupling with a hermitian kernel such that the Green function follows the equation
\begin{equation} \frac{\partial}{\partial Y} {\cal G}(Y,t_1,t_2) 
\ = \ 
  \int  \frac{dt}{\sqrt{\bar{\beta_0} t_1}}  \frac{ {\cal K}(t_1,t)}{\sqrt{\bar{\beta_0} t}}    \, 
{\cal G}(Y,t,t_2). 
\end{equation}
It is convenient to move to Mellin space where the transform can be written in terms of Airy functions~\cite{KLR1,KLR2},  
\begin{equation} {\cal G}_\omega\left( t_1,t_2\right) \ = \ 
 \frac{\pi}{4} 
\frac{\sqrt{t_1t_2}}{\omega^{1/3}} 
\left( \frac{\bar{\beta_0} }{14 \zeta(3)} \right)^{2/3} 
 Ai\left(z(t_1) \right) Bi\left(z(t_2) \right)  \theta\left(t_1-t_2\right)
 \, + \, t_1 \leftrightarrow t_2  \label{mellin-green-1}\end{equation}
with $ z(t) \ \equiv \left(\frac{\bar{\beta_0} \omega}{14\zeta(3) }\right)^{1/3}
 \left(t-\frac{4\ln 2}{\bar{\beta_0} \omega} \right)$.
There is some freedom in the way to present this solution. Keeping in mind that the Green function should vanish when $t_{1,2} \to \infty$, which means that we should  not tamper with the ultraviolet behaviour, it is possible to add to this function any solution of the homogeneous equation with the same UV behaviour. In this way one can replace the Airy function  $Bi(z)$ by $ \overline{Bi}(z) \ \equiv Bi(z) + c(\omega) Ai(z)$. With $c(\omega)$ being 
\begin{equation} c(\omega) \ = \ \cot\left(\eta- \frac{2}{3} \sqrt{\frac{\bar{\beta_0} \omega}{14\zeta(3)}} \left(
 \frac{4\ln 2}{\bar{\beta_0}\omega}-t_0\right)^{3/2}  \right),  
\label{phase} 
 \end{equation}
an infinite number of Regge poles in  the complex $\omega$ plane appear when the argument of this function is 
$- n \pi$. This fixes the phase at the IR fixed point $t_0$ to be $\frac{\pi}{4}+\eta$. 
In principle we can only set the value of 
 $\eta$ using some non-perturbative properties of QCD or, from a more practical point of view, by trying to fit the experimental data. Either way, the interplay between both approaches should teach us some valuable lessons about the confinement region. 
 
\begin{figure}
\vspace{-.3cm}
\begin{center}
\includegraphics[width=7.0cm]{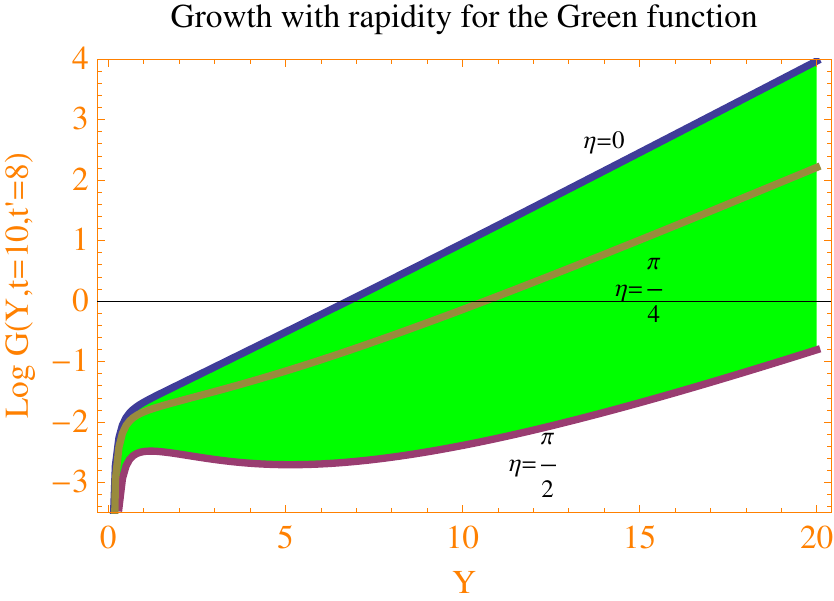}  \includegraphics[width=7.0cm]{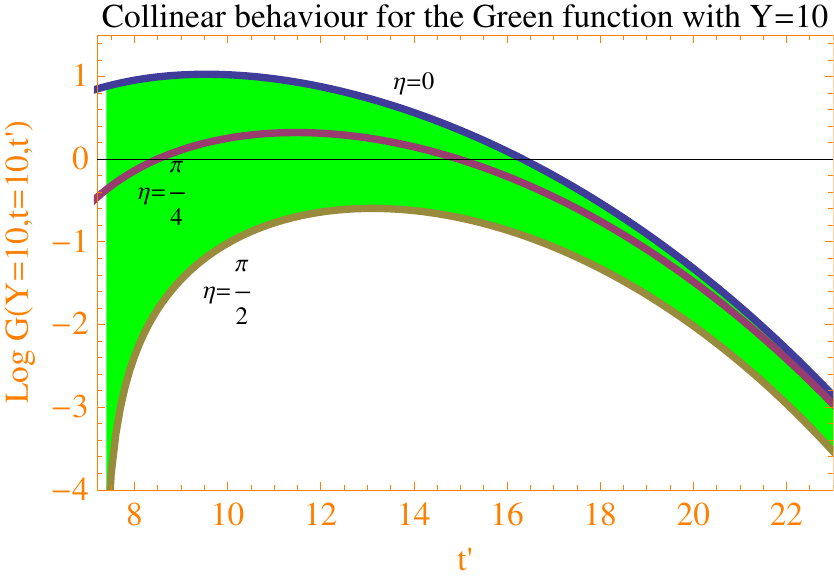}  
\end{center} 
\vspace{-.6cm}
\caption{Effect of the non-perturbative phase $\eta$ on the gluon Green function.}
\label{GGFplot}
\end{figure}

\begin{figure}
\vspace{-.1cm}
\begin{center}
 \includegraphics[width=7.0cm]{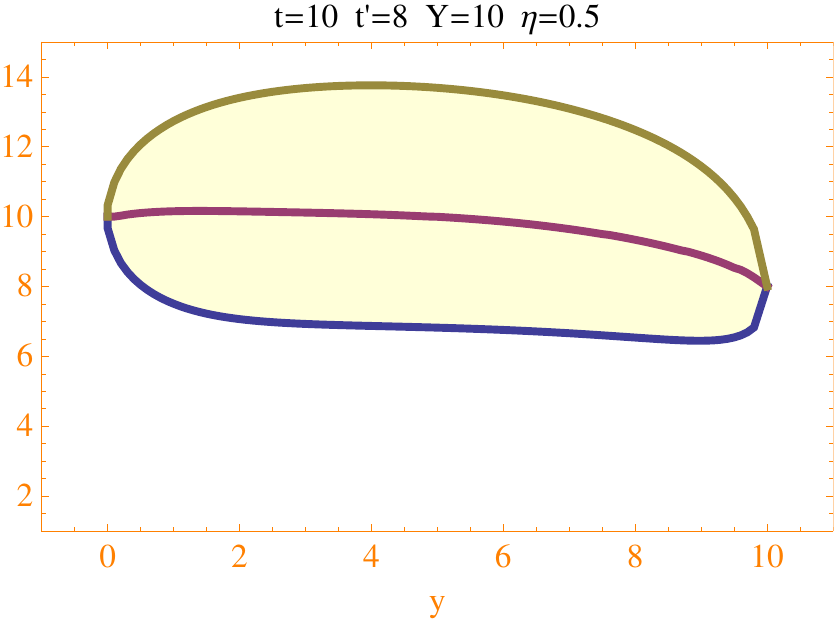}  \, \, \includegraphics[width=7.0cm]{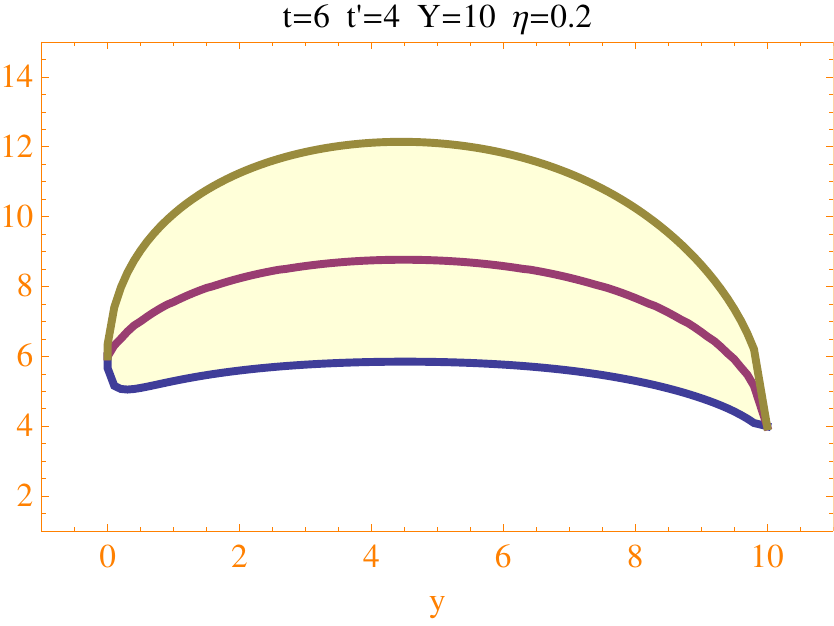}     
\end{center} 
\vspace{-.6cm}
\caption{Effect of the non-perturbative phase on the diffusion profile. }
\label{CigarsPositiveEta}
\end{figure}

Using a quadratic approximation for the kernel simplifies the calculations while keeping the main physical features of the problem. We numerically inverted the Mellin transform and  found that the choice for  $\eta$ affects the phase in Eq.~(\ref{phase}). For example, increasing $\eta$ reduces the initial rise with energy of the solution (LHS plot in Fig.~\ref{GGFplot}).
The collinear behaviour is also influenced, specially at small $t$ (RHS plot in Fig.~\ref{GGFplot}). This is natural since $\eta$ has been introduced in order to modify only the IR sector of the theory. The $t$-profile contains fundamental information about the discrete Pomeron approach since it allows for the study of  the diffusion in transverse scales. This is shown in Fig.~\ref{CigarsPositiveEta}.
For two large and similar external scales (left) the gluon evolution is governed by a  IR/UV symmetric diffusion. This changes when the external scales are smaller (RHS plot) since we find a strong suppression of the diffusion towards the IR. The mean values of the distributions (central line in both plots of  Fig.~\ref{CigarsPositiveEta}) are pushed towards the UV.   This shows how the IR region in the discrete Pomeron approach has been screened out by the choice of boundary conditions. Since it is difficult to fix the non-perturbative phase from first principles, it will be very interesting to find the best fit to data using this as a free parameter.

\section{Rapidity veto constraint}

In multi-Regge kinematics, one assumes that the cascade of gluons emitted between the two primary gluons have a large relative rapidity. 
Following the  original presentation of this idea by Lipatov in~\cite{LevTalk}, in~\cite{Schmidt} it was shown that a significant reduction in the Green function takes place if one considers contributions to the scattering amplitude  in which emitted gluons have a minimum rapidity gap, $b$, relative to the preceding emitted gluon. In~\cite{FRS} it was proven that the effect of imposing such a restriction reproduces the effect of the collinear corrections to the BFKL equation when $b \approx 2$.  This idea has been applied also to non-linear evolution equations~\cite{Chachamis:2004ab}. The mean distance in rapidity among emissions in the BFKL ladder, including collinear contributions, has been studied using {\tt BFKLex} in~\cite{Chachamis:2015ico}. 

\subsection{Effects on the discrete hard Pomeron}

The discrete BFKL spectrum, which we have discussed above, is sensitive to the 
introduction of this lower cut-off in the relative rapidities of the emitted particles.  The eigenvalues associated to each of the discrete eigenfunctions decrease with $b$. This effect is larger on the lowest eigenfunctions. This limits the growth with energy of cross sections and introduces a fast suppression of the regions with small $p_T$.

To be more precise, now we do not use the quadratic approximation for the eigenvalues, $\chi(\nu)$, to write, in the semi-classical approximation, the eigenfunctions of the kernel with running coupling, with eigenvalue $\omega$, in the form 
\beq f_\omega(t) \ = \ \frac{|z_\omega(t)|^{1/4}}{\sqrt{\alphabar(t) \chi^\prime\left(\nu_\omega(t) \right)}}
 Ai\left(z_\omega(t)\right), \eeq
where $\nu_\omega(t) \ = \ \chi^{-1}(\betabar\omega t)$, and $z_\omega(t) \ = \ - \left( \frac{3}{2} \int_t^{4\ln2/\betabar\omega}dt^\prime  \nu_\omega(t^\prime)  \right)^{2/3}$. As explained before, a fixed phase for a negative value of $t$ generates discrete eigenfunctions $f_{\omega_n}(t)$ with  $n$ zeroes. The Green function is now
 \beq {\cal G}(Y,t,t^\prime) \ = \ \sum_n f_{\omega_n}(t) f_{\omega_n}^*(t^\prime) e^{\omega_n Y}. \eeq
A rapidity gap veto  is then implemented in the form 
\beq {\cal G}(Y,t,t^\prime) \ = \ \int_{\cal C} \frac{d\omega}{2\pi i} e^{\omega(Y-b)}   \sum_n \frac{f_{\omega_n}(t) f^*_{\omega_n}(t^\prime)}{\omega-e^{-b \, \omega} \omega_n }. \label{invmell} \eeq
There is a pole  at  $\omega =  \frac{W(b \, \omega_n)}{b}  \equiv \overline{\omega}_n$, where $W(x)$ is the Lambert $W$-function. Hence
\beq 
{\cal G}(Y,t,t^\prime) \ = \    \sum_n  e^{\overline{\omega}_n(Y-b)}
\frac{f_{\omega_n}(t) f^*_{\omega_n}(t^\prime)}{1+b \, \overline{\omega}_n }. \label{green2} 
\eeq

\begin{figure}
\centerline{\includegraphics[width=11.0cm]{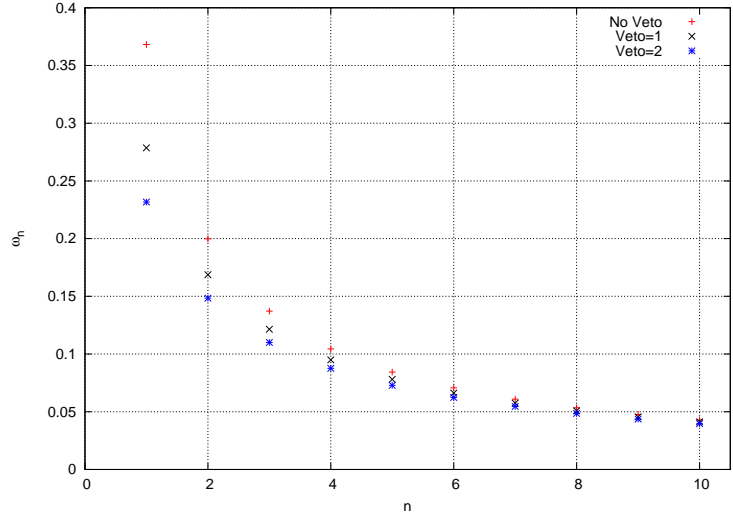}} 
\vspace{-1.cm}
\caption{Effect of introducing a rapidity veto in the effective eigenvalues with $b=1$ (black crosses) and  $b=2$ (blue stars). }
\label{fig2}
\end{figure}
The gap veto damps the solution in two ways. First, with the shift $Y \to Y-b$. Second, with the replacement $\omega_n \to \overline{\omega}_n$ (see their relative size in Fig.~\ref{fig2}).

\subsection{Inclusive dijet hadroproduction}

In~\cite{Caporale:2018qnm} we studied ratios of azimuthal angle distributions in MN jets with a rapidity veto constraint. As we have seen, the asymptotic growth with $Y$ considering only the Green function needs a veto of order 2 to generate positive cross sections. From a phenomenological point of view, we found~\cite{Caporale:2018qnm} the values of the veto which allow for a good fit of angular distributions at the LHC in a realistic (non asymptotic) set-up. 

As we have mentioned, in collisions of two hadrons MN jets have a final state with two tagged jets well separated in rapidity:
\begin{equation} \label{eq:reaction}
p\left(p_A \right) + p\left(p_B\right) \rightarrow J_A \left(k_A\right) + J_B  \left(k_B \right) + X\, .
\end{equation}
The kinematical configuration is $s \equiv \left(p_A + p_B \right)^2 \gg Q^2 \sim \vec{k}_A^2 \sim  \vec{k}_B^2 \gg \Lambda_{QCD}^2$, where $Q$ is a hard transverse scale.  In Sudakov variables,
\begin{equation}\label{eq:SudDecomposition}
k_A=x_{J_A} p_A + \frac{\vec{k}_A^2}{x_{J_A} s} p_B + k_{A,\perp}\, \, , \, \,
k_B=x_{J_B} p_B + \frac{\vec{k}_B^2}{x_{J_B} s} p_A + k_{B,\perp} \, \, , \, \,  k_{A,B,\perp}^2=-\vec{k}_{A,B}^2\, ,
\end{equation}
with $x_{J_{A,B}}$ being the longitudinal momentum fractions of the jets.
The rapidities $y_{A,B}$ of the two tagged jets are 
\begin{equation}\label{eq:rapidities}
y_A=\frac{1}{2}\log{\left( \frac{x_{J_A}^2 s}{\vec{k}_A^2}\right)}, \, \,
y_B=-\frac{1}{2}\log{\left( \frac{x_{J_B}^2 s}{\vec{k}_B^2}\right)}, \, \, 
Y \equiv y_A-y_B=\log{\frac{x_{J_A}x_{J_B}s}{|\vec{k}_A| |\vec{k}_B|}}\,.
\end{equation}
Being a semi-hard process, we need to combine both collinear factorization and BFKL dynamics. We write the cross section as a convolution of the parton distribution functions (PDFs) $f_i \left(x,\mu_F \right)$ and the partonic cross section $\hat{\sigma}$,
\begin{equation}\label{eq:collfactorization}
\frac{\de{}{\sigma} \left( s \right)}{\de{}{y_A}\de{}{y_B}\de{2}{\vec{k}_A} \de{2}{\vec{k}_B}} = \sum_{i,j} \int_{0}^{1} f_i \left(x_A,\mu_F \right) f_j \left(x_B,\mu_F \right) \frac{\de{}{\hat{\sigma}_{ij}} \left(x_A x_B s, \mu_F \right)}{\de{}{y_A}\de{}{y_B}\de{2}{\vec{k}_A} \de{2}{\vec{k}_B}}.
\end{equation}
$i,j$ identify the partons $\left( i,j=q,\bar{q},g\right)$,  $\mu_F$ is the factorization scale and $x_{A,B}$ is the longitudinal momentum fractions of the partons. The partonic cross-section factorizes into jet vertices $V$ and the NLL gluon Green function $\varphi$:
\begin{multline}\label{eq:highenergyfactorization}
\frac{\de{}{\hat{\sigma}_{ij}} \left(x_A x_B s, \mu_F \right)}{\de{}{y_A}\de{}{y_B}\de{2}{\vec{k}_A} \de{2}{\vec{k}_B}}=
\frac{x_{J_A}x_{J_B}}{\left( 2\pi \right)^2} 
\int \frac{\de2{\vec{q}_A}}{\vec{q}_{A}^{\; 2}} V_i\left(\vec{q}_A,x_A,s_0,\vec{k}_A,x_{J_A}, \mu_F, \mu_R \right) 
\\ 
\int \frac{\de2{\vec{q}_B}}{\vec{q}_B^{\;2}} V_j\left(-\vec{q}_B,x_B,s_0,\vec{k}_B,x_{J_B}, \mu_F, \mu_R  \right) 
 \int_C \frac{\de{}{{\omega}}}{2\pi i} \left(\frac{x_A x_B s}{s_0} \right)^\omega \varphi_\omega\left( \vec{q}_A,\vec{q}_B\right)\,.
\end{multline}
All poles in $\omega$ are to the left of the integration contour.  An important piece of the computation is the impact factor at NLO, calculated in~\cite{Bartels:2001ge,Bartels:2002yj,Caporale:2011cc} and later in~\cite{Hentschinski:2011tz,Chachamis:2012gh,Chachamis:2012mw,Chachamis:2012cc}. 

The veto modifies the BFKL kernel only beyond LL accuracy. To introduce some energy-momentum conservation, we expand the Green function in a truncated power series in the coupling to limit the total number of emissions.  

\begin{figure}
\begin{center}
\includegraphics[width=7.cm]{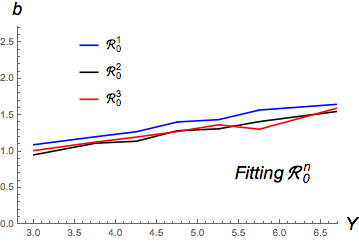}\, \, \, \includegraphics[width=7.cm]{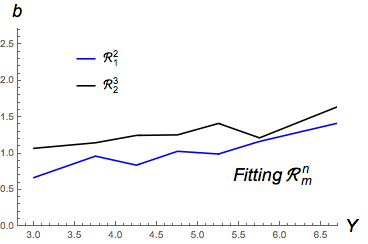} 
\end{center}
\caption{Best-fit values for the rapidity veto $b$ as a function of $Y$ for the ${\cal R}^n_m$ ratios.}
\label{fit}
\end{figure}

We study ratios of azimuthal angle decorrelation~\cite{Vera:2007kn} and compare 
them to CMS~\cite{CMS:2013eda,Khachatryan:2016udy} and ATLAS~\cite{Aad:2011jz} data. The CMS cuts are (for $\sqrt{s}=7 \text{ TeV}$)
\begin{equation}\label{eq:phasespace}
35 \text{ GeV} \leq \norm{\vec{k}_A}, \norm{\vec{k}_B} \leq 60 \textrm{ GeV}, \; \;  \abs{\norm{\vec{k}_A}-\norm{\vec{k}_B}} \geq 2 \text{ GeV}, \; \; 0 \leq y_A, \abs{y_B} \leq 4.7\,. 
\end{equation}
We have calculated the quantities
\begin{equation}\label{eq:coeffphasespace}
C_n (Y,b)= \int \de{2}{\vec{k}_A} \de{2}{\vec{k}_B} \de{}{y_A}\de{}{y_B} \delta(y_A-y_B-Y) \cos{(n \theta)} 
\frac{\de{}{\sigma}}{\de{}{y_A}\de{}{y_B}\de{2}{\vec{k}_A} \de{2}{\vec{k}_B}}
\end{equation}
and the corresponding ratios ${\cal R}^n_m (Y,b) \equiv \frac{C_n (Y,b)}{C_m (Y,b)}$. In Fig.~\ref{fit} the values of the rapidity veto that best fit the LHC data for the correlation functions (left) and the ratios of correlation functions (right) are shown in the $3.25 \le Y \le 6.5$ region. The optimal value of the veto  grows monotonically with $Y$. We have a good understanding of the ratios ${\cal R}^m_n$ including $m=0$ or $n=0$ without having to use the BLM scheme. 
We conclude that a rough cut-off in the rapidity differences among emitted mini-jets in the final state, $b \gtrsim 1$, is enough to obtain a reasonable global description of many different azimuthal angle correlations in dijet cross sections for current LHC data. This value is far from asymptotic estimates. Introducing jet vertices and parton distribution functions at realistic energies reduces it from $b \gtrsim 2$ to half of this value.

\section{\bf Forward Drell--Yan production}

Forward Drell--Yan (f-DY) production at the LHC was studied in~\cite{Brzeminski:2016lwh}. In~\cite{Celiberto:2018muu} we improved their results  by using the unintegrated gluon density of~\cite{Hentschinski:2012kr,Hentschinski:2013id}, calculated at NLL with collinear corrections. We found a good description of the f-DY cross section dependence on the invariant mass of the lepton pair both for LHCb and ATLAS data. This has been proposed~\cite{DYbunch} as an interesting test of the unintegrated gluon density at  small $x$~\cite{Anderson:09/10jca}. 

We showed that it is possible to describe HERA $F_2$ and $F_L$ together with LHC DY data using the NLL BFKL formalism. Our results  should be compared to~\cite{GolecBiernat:2010de,Ducati:2013cga,Schafer:2016qmk,Ducloue:2017zfd}. We studied the production of a lepton-antilepton pair, $L^+ L^-$, in proton-proton collisions:
\begin{equation}
\label{eq:proc}
 p(P_1) \, + \, p(P_2) \; \to \; L^+(l^+) \, + \, L^-(l^-) \, + \, X.
\end{equation}
At LO this is mediated by a virtual photon $\gamma^* (q)$ or a $Z^0 (q)$ boson, where the vector boson momentum $q = l^+ + l^-$ carries a $q_T$ transverse component. $M^2 \equiv q^2 > 0$ is the lepton pair invariant mass squared.

Angular distributions of the lepton pair are written in terms of four structure functions, $\cal{W}_{[\lambda]}$. 
The DY formula~\cite{Lam:1978pu} factorizes in terms of the lepton pair angular phase space $\drv\vartheta^* \, \drv\varphi^* \equiv \drv\Omega_l^*$ ($\vartheta^*$ and $\varphi^*$ are the polar and azimuthal angles of the lepton momentum vector in the dilepton center-of-mass frame) and the structure functions. The $Z^0$ contribution can be neglected and the differential cross section is
\begin{equation}
\label{eq:dsigma}
 \frac{\drv \sigma}{\drv M \, \drv \Omega_l^* \, \drv x_F \, \drv q_T}
 = \frac{\alpha^2 \, q_T}{(2 \pi \, M)^3}
 \left[
 \left( 1 - \cos^2 \vartheta^* \right) {\cal W}_L
 + \left( 1 + \cos^2 \vartheta^* \right) {\cal W}_T
 \right.
\end{equation}
\[
 \left.
 + \; \left( \sin 2 \vartheta^* \cos \varphi^* \right) {\cal W}_\Delta
 + \left( \sin^2 \vartheta^* \cos 2 \varphi^* \right) {\cal W}_{\Delta \Delta}
 \right]
 \; ,
\]
$x_F$ is the longitudinal momentum fraction from the initial-state hadron carried by $\gamma^*$, $\alpha$ the electromagnetic coupling, ${\cal W}_L$ and ${\cal W}_T$ the structure functions for longitudinally and transversely polarized $\gamma^*$, ${\cal W}_\Delta$ the single-spin-flip structure function, and ${\cal W}_{\Delta \Delta}$ the double-spin-flip one. We choose the frame orientation of  Gottfried--Jackson~\cite{Gottfried:1964nx}. In the forward region the structure functions can be written in the form
\begin{equation}
\label{eq:W-HEF}
 {\cal W}_{[\lambda]}\hspace{-.1cm} = \hspace{-.1cm}\frac{2 \pi}{3} \alpha_s (\mu_R) M^2 \hspace{-.2cm} \int_{x_F}^1 \hspace{-.08cm}
 \frac{\drv z}{z^2} 
 f^* \left( \frac{x_F}{z}, \mu_F \right)\hspace{-.2cm}
 \int \hspace{-.15cm} \frac{\drv \kappa_T  \drv \phi_{\kappa_T}}{\left( \kappa_T^2 \right)^2}  \, {\cal G} (x_g, \kappa_T^2) \, \Phi_{[\lambda]} (q_T, \vec{\kappa}_T, z).
\end{equation}
$z$ is the longitudinal momentum fraction of the initial-state quark carried by the virtual photon, $\kappa_T \equiv |\vec{\kappa_T}|$ and  $q_T \equiv |\vec{q_T}|$. The collinear parton distribution, $f^* \left( x, \mu_F \right) = \sum_{r} f_r \left( x, \mu_F \right)$ accounts for incident u, d, s, c and b quarks and antiquarks.

The unintegrated gluon distribution, ${\cal G} (x_g, \kappa_T^2)$, generates the small-$x$ gluon evolution and the f-DY impact factors, $\Phi_{[\lambda]} (q_T, \vec{\kappa}_T, z)$, introduce the $\gamma^* \, \to \, L^+ L^-$ transition. $x_g$ is the gluon longitudinal momentum fraction. 

We took the model for the coupling of the Green function to the proton 
derived in~\cite{Hentschinski:2012kr,Hentschinski:2013id}, also been used in single-bottom quark production~\cite{Chachamis:2015ona}, $J/\Psi$ and $\Upsilon$ photoprotoduction~\cite{Bautista:2016xnp} and $\rho$-meson leptoproduction at HERA~\cite{Bolognino:2018rhb}. This Green function includes the collinear corrections resummed in the form of a $J_1$-Bessel function in~\cite{Vera:2005jt}.

The dependence of the cross-section on the dilepton invariant mass $M$ is
\begin{equation}
\label{eq:sigma}
 \frac{\drv \sigma (M)}{\drv M} = \int \drv \Omega_l^* \int \drv x_F \int \drv q_T \; \frac{\drv \sigma}{\drv M \, \drv \Omega_l^* \, \drv x_F \, \drv q_T} \; .
\end{equation}
In terms of $
 l_T^\pm \equiv \lvert \vec{l}_T^\pm  \rvert = \left(l^\pm \right)^2 - \left( l^{z,\pm} \right)^2$, $\eta^\pm = \textrm{arctanh}\frac{l^{z,\pm}}{l^\pm}$, the LHCb kinematical cuts~\cite{LHCb:2012fja} (with $ 5.5 \mbox{ GeV} < M < 120 \mbox{ GeV}$) are
\begin{equation}
\label{eq:LHCb_cuts}
 2 < \eta^\pm < 4.5
 \; , \quad
 l^\pm > 10 \mbox{ GeV}
 \; , \quad
 \left\{
  \begin{array}{l}
   l^\pm_T > 3 \mbox{ GeV} \hspace{0.55cm} \mbox{if} \hspace{0.35cm} M \le 40 \mbox{ GeV} \\
   l^\pm_T > 15 \mbox{ GeV} \hspace{0.35cm} \mbox{if} \hspace{0.35cm} M > 40 \mbox{ GeV}
  \end{array}
 \right.
\end{equation}
In~\cite{Celiberto:2018muu} predictions are also given for the ATLAS experimental cuts~\cite{Piccaro:2012kzm,Aad:2014qja}. 

\begin{figure}
\centering
 \includegraphics[scale=0.4,clip]{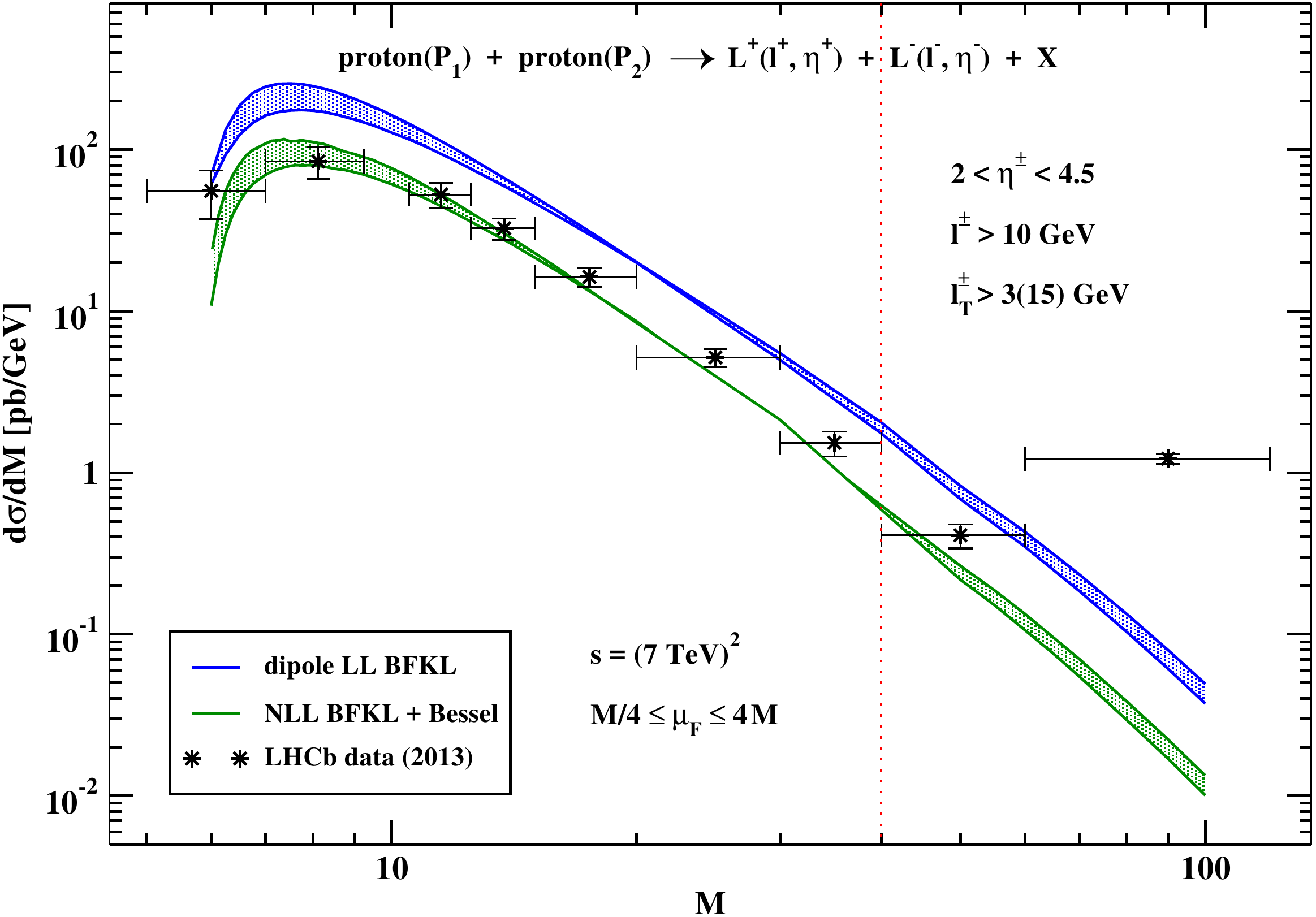}
\vspace{-.3cm}
\caption{DY cross section as a function of dilepton invariant mass $M$. Uncertainty bands account for factorization scale shifts $M/4 \le \mu_F \le 4M$.}
\label{fig:fDY-NLO-LHCb}
\end{figure}

Fig.~\ref{fig:fDY-NLO-LHCb} shows that our calculations reproduce well the LHCb data with the same set of parameters that also fit the HERA data.
This is a positive result for the BFKL approach since this global description is a mandatory feature of the framework. The same DY data can, however, be also  described by fixed order calculations~\cite{Aad:2014qja} or studies which include saturation corrections~\cite{Brzeminski:2016lwh}. In order to disentangle different approaches, future LHC data for f-DY pair production~\cite{N.Cartiglia:2015gve} will be very important. \\
\\
The research here presented has been supported by the Spanish Research Agency (Agencia Estatal de Investigaci\'on) through the grant IFT Centro de Excelencia Severo Ochoa SEV-2016-0597, and the Spanish Government grants FPA2015-65480-P, FPA2016-78022-P.

\end{document}